\documentclass[twocolumn,prl,showpacs,aps,floatfix]{revtex4}

\usepackage{amssymb}
\usepackage{bm}
\usepackage[usenames]{color}
\usepackage{graphicx}

\newcommand{\be}{\begin{equation}}
\newcommand{\ee}{\end{equation}}
\newcommand{\br}{\begin{eqnarray}}
\newcommand{\er}{\end{eqnarray}}

\begin{document}

\title{Complete State Reconstruction of a Two-Mode
Gaussian State via Local Operations and Classical Communication}
\author{Gustavo Rigolin}
\email{rigolin@ifi.unicamp.br}
\author{Marcos C. de Oliveira}
\email{marcos@ifi.unicamp.br}
\affiliation{ Instituto de F\'\i
sica ``Gleb Wataghin'', Universidade Estadual de Campinas,
Unicamp, 13083-970, Campinas, S\~ao Paulo, Brasil}


\begin{abstract}
We propose a strictly local protocol completely equivalent to global 
quantum state reconstruction for a bipartite 
system. We show that the joint
density matrix of an arbitrary two-mode Gaussian state, entangled
or not, is obtained via local operations and classical
communication only. In contrast to previous proposals, simultaneous
homodyne measurements (HM) on both modes are replaced by local homodyne 
detections and a set of local projective measurements. 
\end{abstract}



\maketitle


The feasibility of a quantum information task is related to the 
reduced or absence of non-local resources needed to
its implementation and is an important asset for quantum
communication purposes \cite{braunstein}, setting the limit for its 
widespread use. However, Quantum State Tomography (QST) \cite{vogel,raymer}, 
a key tool in quantum information, is performed mostly 
through non-local operations. 
QST is a complete state reconstruction scheme implemented through a set of 
measurements over an ensemble of identical 
quantum systems. For qubit systems it corresponds to the 
determination of all the Stokes parameters \cite{QSE}. 
For Gaussian continuous variable (CV) systems, 
as given by quantized electromagnetic field modes, 
it stands on a set of joint quadrature measurements, from which the joint 
density matrix $\rho$ is reconstructed.
 Thus for Gaussian states,
QST is equivalent to the measurement of 
global covariance matrices of the modes.
For a two-mode Gaussian state most QST protocols to date either 
require simultaneous HM on both modes
\cite{vasilyev,babichev,bowen04}, 
with an exquisite control of both local oscillators 
(LO) phases, or require previous non-local operations
on the modes to achieve a complete state reconstruction
\cite{walls}. It is desirable, therefore, the construction
of a QST protocol that does not require any non-local operation and no phase-locking. In other words, a process which is operationally  
equivalent to QST, but without unnecessary non-local resources to its 
implementation.

%

In this paper we
show how one can reconstruct
the whole density matrix $\rho_{12}$ of an arbitrary two-mode
Gaussian state via local operations and classical communication
(LOCC) only. Since simultaneous HM of the two modes
\cite{vasilyev,babichev} are not required, there is no need for 
constrained control of the LO's phases, thus
increasing the overall efficiency of the protocol, and also
reducing the computational post-processing of data (See Ref. \cite{footnote2} 
for an interesting single homodyning alternative scheme). 
 Instead, a set of local parity 
and vacuum projections plus 
local squeezing are required.
Our protocol is built
basically on three premises: \textit{(i)} Alice and
Bob can implement independent single mode local QST, certifying that they have 
a Gaussian state. Actually, after confirming (or being
informed previously) that one deals with a Gaussian state, only HM's 
of the variances of the modes will suffice. \textit{(ii)} Both Alice and Bob 
are able to implement local squeezing and a local rotation on the
quadratures of their modes. \textit{(iii)} Bob
(or Alice) can make two types of local measurements: even/odd
parity projections and vacuum projections of his (or her) mode.

A bipartite
two-mode Gaussian state $\rho_{12}$ is completely described
\cite{englert,review-adesso} by its Gaussian characteristic
function $C({\bm
z})=\mbox{Tr}(D(z)\rho_{12})=e^{-\frac12{\bm{z}^\dagger}{\bf
V}{\bm{z}} }$, where $\bm{z}^\dagger=\left(z_1^*, z_1, z_2^*,
z_2\right)$ are complex numbers,
$D(z)=e^{-\mathbf{z^\dagger}\mathbf{E}\mathbf{v}}$ is the
displacement operator, with ${\bf
E}=\mbox{diag}({\bm{Z}},{\bm{Z}})$, ${\bm{Z}}=\mbox{diag}(1,-1)$,
and
$\mathbf{v}=(v_1,v_2,v_3,v_4)^{T}=(a_1,a_1^\dagger,a_2,a_2^\dagger)^{T}$
the annihilation and creation operators of modes 1 and 2,
respectively. $T$ is the transposition, so that
$\mathbf{v}$ is a column vector, and we have assumed all the
first order moments to be null \cite{footnote1}. The covariance
matrix \textbf{V} describing all the second order moments
$V_{ij}=(-1)^{i+j}\langle v_i v_j^\dagger + v_j^\dagger v_i
\rangle/2$ is given by
\begin{equation} \textbf{V}=\left(
\begin{array}{cc}
\textbf{V}_1 & \textbf{C} \\
\textbf{C}^\dagger & \textbf{V}_2 \end{array} \right)=\left(
\begin{array}{cccc}
n_1 & m_1 & m_s & m_c \\
m_1^* & n_1 & m_c^* & m_s^* \\
m_s^* & m_c & n_2 & m_2 \\
m_c^* & m_s & m_2^* & n_2 \\ \end{array} \right).
\label{covariancia}
\end{equation}
Here $\textbf{V}_1$ and $\textbf{V}_2$ are the local covariance
matrices of modes $1$ and $2$, respectively, giving the local
properties of the two modes while \textbf{C} is the correlation
between them. Finally, in addition to being positive semidefinite,
$\textbf{V}\geq\mathbf{0}$, a physical Gaussian state must satisfy
the generalized uncertainty principle, $\textbf{V}+\frac 1
2\textbf{E}\geq\mathbf{0}$ \cite{englert}.

The main goal of Alice and Bob is to obtain via LOCC the matrix
$\mathbf{V}$. Therefore, the first logical and trivial step
consists in the measurement of $\mathbf{V_1}$ and $\mathbf{V_2}$
by Alice and Bob, respectively. These two covariance matrices are
locally obtained via any standard single mode HM
technique (or local QST). Up to now no classical communication is
needed and only after finishing this task Bob (Alice) informs
Alice (Bob) of his (her) result. It is worth noting that we assume
Alice and Bob have at their disposal a trustful source, in the
sense that it produces as many as needed identical copies of the
two-mode Gaussian state.

\begin{figure}[!ht]
\includegraphics[width=8cm]{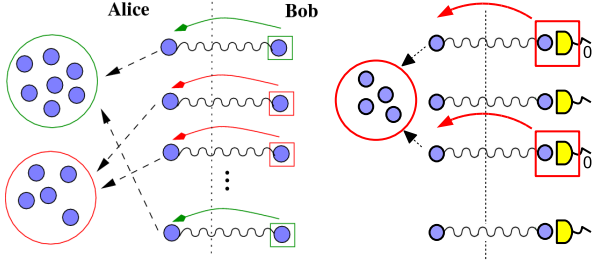}
\caption{\label{fig-groups}(Color online) \textit{Left:} Alice separates her
copies in two groups conditioned on an even (green) or odd (red)
parity result obtained by Bob. 
\textit{Right:} Alice selects copies corresponding to 
Bob's no-photon results.}
\end{figure}

The next non trivial step is the determination of $\mathbf{C}$. 
To achieve such a goal, Alice and Bob need to work
collaboratively \cite{haruna}. First, on a subensemble of the
copies, Bob implements parity measurements on his mode and informs
Alice the respective outcomes for each copy, i.e., even parity
(even number of photons) or odd parity (odd number of photons).
With this information Alice separates her copies in two distinct
groups, the even (e) and the odd (o) ones \cite{haruna},
as depicted in Fig.~\ref{fig-groups}.
Alice's even group can be described by the non-normalized density
matrix
$\rho_{1}^{e}=\mbox{Tr}_2\left\{P^{e}_2\rho_{12}P^{e}_2\right\} =
\sum_{n=0}^{\infty}\,_2\langle 2n | \rho_{12} | 2n \rangle_2$,
where $P^{e}_2=\mathcal{I}_1\otimes
\sum_{n=0}^{\infty}|2n\rangle_2\,_2\langle 2n|$, $\mathcal{I}_1$
the identity operator, and $|n\rangle_2$ the $n$-th Fock
state for mode $2$. Using a similar notation, Alice's odd group is
given as
$\rho_{1}^{o}=\mbox{Tr}_2\left\{P^{o}_2\rho_{12}P^{o}_2\right\} =
\sum_{n=0}^{\infty}\,_2\langle 2n+1 | \rho_{12} | 2n+1 \rangle_2$.
But one can show that \cite{haruna2}
%
$ \sigma_1 = 2\sqrt{\mbox{det}(\mathbf{V}_2)} (\rho_{1}^{e}
\,-\, \rho_{1}^{o}) =\int\mathrm{d}\mathbf{z_1}
\mathrm{e}^{\mathbf{z_1^\dagger}\mathbf{Z}\mathbf{a}_1}\mathrm{e}^{-\frac{1}{2}
\mathbf{z_1^\dagger}\mathbf{\Gamma_1}\mathbf{z_1}}, $
%
where $\mathrm{d}\mathbf{z_1} = (1/\pi)
\mathrm{d}\mbox{Re}(z_1)\mathrm{d}\mbox{Im}(z_1)$ and
$\mathbf{\Gamma_1}$ is the Schur complement \cite{horn} of
$\mathbf{V_2}$:
\begin{equation} {\bf
\Gamma}_1=\textbf{V}_1-\textbf{C}\textbf{V}_2^{-1}\textbf{C}^\dagger
= \left(
\begin{array}{cc}
\eta_1 & \mu_1 \\
\mu_1^* & \eta_1\end{array} \right). \label{schur_rel1}
\end{equation}
However, any one-mode Gaussian operator can be written as
 $\int\mathrm{d}\mathbf{z_1}
\mathrm{e}^{\mathbf{z_1^\dagger}\mathbf{Z}\mathbf{a}_1}\mathrm{e}^{-\frac{1}{2}
\mathbf{z_1^\dagger}\mathbf{\Gamma}_1\mathbf{z_1}}$, being
$\mathbf{\Gamma}_1$ its covariance matrix \cite{englert}.
Therefore, $\sigma_1$ is a Gaussian operator whose covariance
matrix elements are
%
$ \eta_1=2\sqrt{\mbox{det}(\mathbf{V}_2)}(\langle a_1^\dagger
a_1\rangle_{e}-\langle a_1^\dagger a_1\rangle_{o}) $ and $
\mu_1=2\sqrt{\mbox{det}(\mathbf{V}_2)}(\langle
a_1^2\rangle_{e}-\langle a_1^2\rangle_{o}),
$
%
where $\langle \cdot \rangle_e = \mbox{Tr}(\cdot \rho_1^e)$ and
$\langle \cdot \rangle_o=\mbox{Tr}(\cdot \rho_1^o)$. Summing up,
$\mathbf{\Gamma}_1$ can be obtained with the knowledge of
$\mbox{det}(\mathbf{V}_2)$ and the second moments of $\rho_1^e$
and $\rho_1^o$, all of which determined via LOCC \cite{nonGaussian}.
Defining $\gamma = (n_1-\eta_1)\left(n_2^2-\vert
m_2\vert^2\right)$ and $ \delta = (m_1-\mu_1)\left(n_2^2-\vert
m_2\vert^2 \right)$, Eq. (\ref{schur_rel1}) gives two independent
equations,
which alone cannot give $m_s$ and $m_c$ unequivocally:
\begin{eqnarray}
\gamma &=& n_2 \left(\vert m_c\vert^2+\vert m_s\vert^2\right)
-2\mbox{Re}(m_2m_sm_c^*), \label{eq-gama}\\
\delta &=& 2 n_2 m_s m_c - m_2^*m_c^2-m_2m_s^2. \label{eq-delta}
\end{eqnarray}
%

A unique solution though can be obtained if we consider an
additional subensemble on which Bob performs another kind of 
projective measurement. The results of 
this measurement are communicated to Alice who build a local covariance 
matrix that is related to the original one through the Schur complement 
structure, similar to (\ref{schur_rel1}). In the present case we consider 
the simplest choice, i.e., Bob is able to perform a vacuum state
projection on his copies: photon-number measurements with no outcome. 
For each measurement, 
Bob informs Alice to which copies a
no-photon result ($\rho_2 \rightarrow |0\rangle_2\,_2\langle 0|$)
occurred. Alice, then, proceeds in a similar fashion as before but 
considering only the
vacuum projected subensemble (right of Fig.~\ref{fig-groups}), 
described by the
density matrix $\rho_1^{vac} = \mbox{Tr}_2(|0\rangle_2\,_2\langle
0|\rho_{12})/\mbox{Tr}_{12}(|0\rangle_2\,_2\langle 0|\rho_{12})$. 
One can show that \cite{haruna2} $\rho_{1}^{vac} =
\int\mathrm{d}\mathbf{z_1}
\mathrm{e}^{\mathbf{z_1^\dagger}\mathbf{Z}\mathbf{a}_1}\mathrm{e}^{-\frac{1}{2}
\mathbf{z_1^\dagger}\mathbf{\Pi_1}\mathbf{z_1}}$, 
where
\be {\bf
\Pi}_1=\textbf{V}_1-\textbf{C}\left(\textbf{V}_2+\frac{1}{2}
\textbf{I}\right)^{-1}
\textbf{C}^\dagger =\left(
\begin{array}{cc}
\xi_1 & \nu_1 \\
\nu_1^* & \xi_1 \end{array} \right),  \label{vacuo} \ee
with $\textbf{I}$ the identity matrix of dimension two. Here, 
$\xi_1=\langle a_1^\dagger a_1\rangle_{vac}$ and  $\nu_1=\langle
a_1^2\rangle_{vac}$, where $\langle \cdot
\rangle_{vac}=\mbox{Tr}(\cdot\rho_1^{vac})$. Explicitly,
Eq.~(\ref{vacuo}) gives us two more equations,
\begin{eqnarray}
\alpha &=& \left(n_2+\frac{1}{2}\right) \left(\vert
m_c\vert^2+\vert m_s\vert^2\right)
-2\mbox{Re}(m_2m_sm_c^*), \label{eq-alfa}\\
\beta &=& 2 \left(n_2+\frac{1}{2}\right) m_s m_c -
m_2^*m_c^2-m_2m_s^2, \label{eq-beta}
\end{eqnarray}
in which $\alpha = (n_1-\xi_1)((n_2+1/2)^2-\vert m_2\vert^2)$ and
$\beta = (m_1-\nu_1)((n_2+1/2 )^2-\vert m_2\vert^2 )$. It is worth
to stress that $\alpha$ and $\beta$, as well as $\gamma$ and
$\delta$, are functions of parameters locally obtained by Alice
and Bob.
In order to solve Eqs.~(\ref{eq-gama}), (\ref{eq-delta}),
(\ref{eq-alfa}), and (\ref{eq-beta}) for $m_c$ and $m_s$ we write
$ m_j = |m_j|e^{i\theta_j},
$
where $j=1,2,c,s$ and $\theta_j$ is real. In this notation, our
task is to determine $\theta_c$, $\theta_s$, $|m_c|$, and $|m_s|$ 
with the aid of $\mathbf{\Gamma_1}$ and $\mathbf{\Pi_1}$, 
obtained by Alice via LOCC. 
The other quantities, $n_1$, $n_2$, $|m_1|$, $|m_2|$,
$\theta_1$, and $\theta_2$ are easily obtained via local HM of
modes $1$ and $2$. 

\textit{(i) Determination of $\theta_c$ and
$\theta_s$}. Subtracting Eq.~(\ref{eq-delta}) from (\ref{eq-beta})
we have
%
$
\beta - \delta = m_s m_c,
$
%
which gives
\begin{eqnarray}
|m_s m_c| &=& |\beta - \delta|, \label{produto-dos-modulos}\\
\theta_s + \theta_c & = & Arg(\beta - \delta)
\label{soma-das-fases},
\end{eqnarray}
where $Arg(z)$ is the phase of the complex number $z$. By the same
token, subtracting Eq.~(\ref{eq-gama}) from (\ref{eq-alfa}) we
have
\begin{equation}
|m_c|^2 + |m_s|^2 = 2(\alpha - \gamma). \label{soma-dos-quadrados}
\end{equation}
Inserting Eqs.~(\ref{produto-dos-modulos}) and
(\ref{soma-dos-quadrados}) into (\ref{eq-alfa}) we get,
%
$
\alpha = (2n_2 + 1)(\alpha - \gamma) - 2 |m_2(\beta -
\delta)|\cos(\theta_2 + \theta_s - \theta_c).
$
%
We could have used Eq.~(\ref{eq-gama}) as well. Solving,
then, for $\theta_s - \theta_c$ we obtain,
\begin{equation}
\theta_s - \theta_c = \cos^{\!-1}\!\left[(\alpha n_2 - \gamma(n_2
+ 1/2))/|m_2(\beta - \delta)|\right] - \theta_2.
\label{diferenca-das-fases}
\end{equation}
Eqs.~(\ref{soma-das-fases}) and (\ref{diferenca-das-fases}) can be
easily solved to give $\theta_c$ and $\theta_s$, the phases of
$m_c$ and $m_s$. It is worth mentioning that
Eq.~(\ref{diferenca-das-fases}) is only valid when $|(\beta -
\delta) m_2| \neq 0$. Later we show how to overcome this
limitation.

\textit{(ii) Determination of $|m_c|$ and
$|m_s|$.} From Eq.~(\ref{soma-dos-quadrados}) we note that if
we had $|m_c|^2-|m_s|^2$ the problem would be solved. Manipulating
the real and imaginary parts of Eq.~(\ref{eq-beta}) we get
%
\begin{equation}
|m_c|^2 - |m_s|^2 = |\beta|\sin(\theta_\beta - \theta_c -
\theta_s)/[|m_2|\sin(\theta_2 - \theta_c + \theta_s)].
\label{diferenca-dos-quadrados}
\end{equation}
Here $\theta_\beta$ is the phase of $\beta$.
Eqs.~(\ref{soma-dos-quadrados}) and
(\ref{diferenca-dos-quadrados}) can be directly solved to give
$|m_c|$ and $|m_s|$, the moduli of $m_c$ and $m_s$. 
Eq.~(\ref{diferenca-dos-quadrados}) is only valid for
$|m_2\sin(\theta_2 - \theta_c + \theta_s)| \neq 0$.  Thus, all the 
covariance matrix elements can be locally reconstructed with a set of 
appropriate measurements and classical communication, establishing 
the following important connection to Gaussian QST:
%
\textit{Global QST is completely equivalent to local covariance matrix HM,
local parity 
and vacuum state projections, 
and 
classical communication.} 
This is our central result and in the rest of this Letter we show how the 
necessary conditions $|(\beta -\delta) m_2| \neq 0$ and 
$|m_2\sin(\theta_2 - \theta_c + \theta_s)|\neq 0$ can always be obtained by 
the addition of local squeezing \cite{localSqueezing}.

\textit{(iii) Overcoming $|(\beta -
\delta) m_2| = 0$ or $|m_2\sin(\theta_2 - \theta_c + \theta_s)| = 0$.}
To properly solve these problems we must know which
quantity is zero. The simplest check is implemented
when Bob reconstructs $\mathbf{V_2}$, which allows him to know
if $m_2=0$. Alice and Bob can also discover if
$m_c=m_s=0$ (implying $\beta -\delta=m_sm_c=0$) by testing if
$\mathbf{V_1}=\mathbf{\Gamma_1}=\mathbf{\Pi_1}$, 
since the absence of correlation ($\mathbf{C}=0$) between the
modes cannot change what the parties measure locally (See
Eqs.~(\ref{schur_rel1}) and (\ref{vacuo})). Also, if
$\mathbf{V_1}\neq \mathbf{\Gamma_1}$ or $\mathbf{V_1}\neq
\mathbf{\Pi_1}$ Alice and Bob are sure that $\mathbf{C}\neq 0$ and
the first non-trivial check sets in. They must discover if either
$m_c=0$ and $m_s\neq 0$, or $m_c\neq0$ and $m_s= 0$, or both
$m_c\neq0$ and $m_c\neq0$.
If either $m_c$ or $m_s$ is zero it is obvious that
$|I_3|=|\mbox{det}(\mathbf{C})|= ||m_s|^2 - |m_c|^2| = |m_s|^2 +
|m_c|^2.$ But one can show \cite{haruna} that
$|I_3|=\sqrt{\mbox{det}(\mathbf{V}_2)\mbox{det}(\mathbf{V}_1 -
\mathbf{\Gamma}_1)}$ and using Eq.~(\ref{soma-dos-quadrados}) we
see that if $|I_3|=2(\alpha- \gamma)$ we know for sure that either
$m_s$ or $m_c$ is zero. If we do not have an equality $m_c\neq0$
and $m_s\neq0$. For our purposes, as we explain below, we do not
need to know which quantity, $m_c$ or $m_s$, is zero \cite{footnoteEnt}. 
Finally, to
discover if $\sin(\theta_2 - \theta_c + \theta_s)=0$ we use
Eq.~(\ref{diferenca-das-fases}) and the phase of $m_2$. Of
course, Eq.~(\ref{diferenca-das-fases}) is only valid if
$|m_2(\beta-\delta)|\neq0$. Therefore, if $|m_2(\beta-\delta)|=0$
we first need to solve this problem in order to test if
$\sin(\theta_2 - \theta_c + \theta_s)=0$.
Since now we know 
which parameter is zero we are
ready to show how Alice and Bob can overcome this situation
allowing them to use Eqs.~(\ref{soma-das-fases}) to
(\ref{diferenca-dos-quadrados}) to obtain $\mathbf{C}$. 
See Tab. \ref{table1} for an overview of the strategies to 
solve these problems. 
\begin{table}[!ht]
\vspace{-.35cm}
\caption{\label{table1} Overview of the general strategies. 
Here $j=2,s,c$.}
\begin{ruledtabular}
\begin{tabular}{lcc}
 &  $m_j=0$ & $\sin(\theta_2 - \theta_c + \theta_s) = 0$ \\ \hline
local squeezing & yes  & yes\\
local quadrature rotation & no & yes 
\end{tabular}
\end{ruledtabular}
\vspace{-.25cm}
\end{table}

If $m_2= 0$ the most general solution \cite{footnoteM2} 
is achieved implementing a
local symplectic transformation (local quadrature squeezing and rotation) on 
mode $2$ \cite{Sim94}, $\mathbf{S} = \mbox{diag}(\mathbf{I}_1,\mathbf{S_2})$,
%
%
where $\mathbf{I}_1$ is a $2\times 2$ identity matrix acting on system 1 and 
$\mathbf{S}_2$ is given as
\begin{equation}
\mathbf{S}_2 = \left(
\begin{array}{cc}
e^{-is_2}\cosh r_2&
\sinh r_2 \\
\sinh r_2&
e^{is_2}\cosh r_2
\end{array}
\right),
\end{equation}
being $s_2$ and $r_2$ real parameters. The new correlation
matrix $\tilde{\mathbf{V}}$ is connected to $\mathbf{V}$ by 
$
\tilde{\mathbf{V}} = \mathbf{S} \mathbf{V} \mathbf{S^{\dagger}}
%
$ \cite{Sim94},
or equivalently for $j=1,2$,%
$\mathbf{\tilde{V}}\!_j = \mathbf{S}_j 
\mathbf{V}\!_j\,\mathbf{S^{\dagger}}\!_j,   \mathbf{\tilde{C}}
= \mathbf{S}_1 \mathbf{C}\mathbf{S^{\dagger}}\!_2.$
%
Applying $\mathbf{S}$ to (\ref{covariancia}), the off-diagonal
term of $\tilde{V}_2$ is $\tilde{m}_2 = \mathrm{e}^{-2 i s_2}m_2\cosh^2r_2
+m_2^*\sinh^2r_2 + \mathrm{e}^{-is_2}n_2\sinh(2r_2).$
%
%
Setting $s_2=0$ and using that $m_2=0$ we have
\begin{equation}
\tilde{m}_2 = n_2\sinh(2r_2), \label{eq-m2til-m2zero}
\end{equation}
i.e., a new covariance matrix with $\tilde{m}_2\neq 0$.
After this operation we can proceed with
the original protocol to reconstruct $\mathbf{\tilde{V}}$, which can be
transformed back to give $\mathbf{V} = \mathbf{S}^{-1} \mathbf{\tilde{V}}
\mathbf{S^{\dagger}}^{-1}$,  with $\mathbf{S}^{-1} = \mbox{diag}\left(
\mathbf{I}_1, \mathbf{S}_2^{-1}\right)$
%
%
%
%
and
\begin{equation}
\mathbf{S}_2^{-1} = \left(
\begin{array}{cc}
e^{is_2}\cosh r_2&
-\sinh r_2 \\
-\sinh r_2& e^{-is_2}\cosh r_2
\end{array}
\right).
\end{equation}

If either $m_s=0$ or $m_c=0$, or equivalently $\beta - \delta =0$,
we can obtain a new matrix 
$\tilde{\mathbf{C}}= \mathbf{S}_1 \mathbf{C}\mathbf{S^{\dagger}}\!_2$ where both
parameters are not zero via a local squeezing operation alone. This leads to
%
\begin{eqnarray}
\tilde{m}_s &=& e^{is_2}m_s\cosh r_2 + m_c\sinh r_2, \label{newms}\\
\tilde{m}_c &=& e^{-is_2}m_c\cosh r_2 + m_s\sinh r_2.
\label{newmc}
\end{eqnarray}
Setting $s_2=0$ in Eqs.~(\ref{newms}) and (\ref{newmc}) we see
that $\tilde{m}_s$ and $\tilde{m}_c$ are combinations of $m_s$ and
$m_c$. Therefore, if $m_s=0$ or $m_c=0$ the new coefficients are
necessary different from zero whenever we apply a local squeezing
operation on mode $2$. As anticipated, we do not need
to know which quantity was originally zero. As before, after this
local transformation we proceed with the original protocol
obtaining $\mathbf{\tilde{V}}$ and then $\mathbf{V}$.
It is worth noting  that when the two
situations occur simultaneously, i.e. $m_2=0$ and $m_s=0$ or
$m_c=0$, the same local squeezing operation solves at once both
problems, as can be seen in Eqs.~(\ref{eq-m2til-m2zero}),
(\ref{newms}), and (\ref{newmc}).

Lastly, after being sure that $|m_2m_cm_s|\neq0$ we can proceed to
test if $\sin(\theta_2 - \theta_c + \theta_s)=0$ using
Eq.~(\ref{diferenca-das-fases}) and the phase of $m_2$, all
quantities locally determined. In case of a positive result, there
exist three possible solutions. The first one is valid when
$m_2\neq m_1$ and is achieved reversing the roles of Alice and Bob
in the protocol, as discussed above. The remaining two
possibilities, and more general, is to locally and unitary
transform mode $1$ or mode $2$ before we implement the protocol,
in the same fashion as before. Therefore, we need to show that
there exists at least one local unitary operation acting on mode
$1$ or mode $2$ that eliminates such a problem.

Let us begin with mode $2$. Applying the symplectic local
transformation $\mathbf{S_2}$ 
 we get, after assuming
that $\sin(\theta_2 - \theta_c + \theta_s) = 0$,
\begin{equation}
\tan(\tilde{\theta}_2 + \tilde{\theta}_s - \tilde{\theta}_c) =
2\mathcal{A}_{\pm}\sin(\theta_2-s_2)\sinh(2r_2)/\mathcal{B}_{\pm},
\label{tan}
\end{equation}
%
%
\begin{eqnarray}
\mathcal{A}_\pm &=& |m_2|(|m_c|^2+|m_s|^2) \mp 2n_2|m_cm_s|, \label{amais}\\
\mathcal{B}_\pm &=& \pm 3|m_2m_cm_s| - n_2(|m_c|^2+|m_s|^2)
\nonumber\\
&& + (\pm|m_2m_cm_s|+n_2(|m_c|^2+|m_s|^2))\cosh(4r_2) \nonumber \\
&& \pm 2 |m_2m_cm_s| \cos(2\theta_2 - 2s_2)\sinh^2(2r_2) \nonumber\\
&& + \mathcal{A}_\mp \cos(\theta_2 - s_2)\sinh(4r_2).
\label{bmais}
\end{eqnarray}
Here $\mathcal{A}_\pm$ ($\mathcal{B}_\pm$) stand for the two
possible values for the cosine, i.e., $\cos(\theta_2 - \theta_c +
\theta_s) = \pm 1$, respectively. From Eqs.~(\ref{tan}) and
(\ref{amais}) we see that a local squeezing alone ($r_2\neq 0$ and
$s_2=0$) on mode $2$ can make $\tan(\tilde{\theta}_2 +
\tilde{\theta}_s - \tilde{\theta}_c) \neq 0$ if $m_2$ is not real
($\theta_2 \neq 0$). However, whenever $m_2$ is real a rotation on
the quadratures ($s_2\neq 0$) is mandatory.
There is one last loophole to fix, namely, the rare instances in
which $\mathcal{A}_+=0$ (note that $\mathcal{A}_-$ is always
different from zero). This is fixed by allowing the other party,
in this case Alice, to implement a local squeezing on mode $1$. As
shown in Eqs.~(\ref{newms}) and (\ref{newmc}) this operation
allows Alice to change at her will the phases of $m_s$ and $m_c$
without altering $\theta_2$, solving completely the last problem.
By the way, this is 
the other possible solution for the
$\sin(\theta_2 - \theta_c + \theta_s) = 0$ case, i.e., 
a local squeezing 
directly on mode $1$.

In summary, we showed a strictly local protocol in which a
two-mode Gaussian state is completely reconstructed without
relying on simultaneous HM or non-local resources. 
Actually, the only resources needed for this protocol are the ability to 
perform single mode HM, local parity and vacuum projective measurements, 
and classical communication. We also showed the complete 
equivalence of this local protocol to QST for Gaussian states. 
This equivalence 
is important for
quantum communication purposes since now we can achieve the same goals of QST 
without non-local resources and simultaneous HM.
The set of local parity measurements required here, however, may restrict the 
implementation of the protocol, apart from instances where this measurement 
can be, in principle, performed \cite{gerry}. 
Finally, this new
protocol raises several interesting problems yet to be solved.
Firstly, it is unknown if a similar local protocol can be devised
for more than two modes and, secondly, if there exist
other optimal sets of measurements, than parity and vacuum projections,
allowing the complete state reconstruction of a two-mode (or
many-mode) Gaussian (or non-Gaussian) state in a simpler local way.

\begin{acknowledgments}
GR and MCO thank FAPESP and CNPq for funding.
\end{acknowledgments}


\begin{thebibliography}{99}
\bibitem{footnote1}  This can be achieved through a local displacement 
operation without changing the states' entanglement content.
\bibitem{nonGaussian} We do not need to manipulate these non-Gaussian states;
we just measure their covariance matrices.
\bibitem{footnoteEnt} We can discover which parameter is zero
if the state is entangled because in this case $I_3<0$ \cite{simon}, 
implying that $m_s=0$ and $m_c\neq0$. To test for entanglement 
we use the Simon separability test \cite{simon}:  we have
entanglement if, and only if, $I_1 I_2 + (1/4 - |I_3|)^2 - I_4 <
(I_1 + I_2)/4,$ with $I_j=\mbox{det}(\mathbf{V}_j)$, $j=1,2$,
$I_4=I_1 I_2 + I_3^2 - I_V$, $I_V=\mbox{det}(\mathbf{V})=
\mbox{det}(\mathbf{V}_2)\mbox{det}(\mathbf{\Gamma}_1);$ 
$\mathbf{C}$ is also readily attained since
$|m_c|^2=\alpha/(n_2+1/2)=\gamma/n_2$ and $\theta_c=(\theta_\beta
+ \theta_2 - \pi)/2 = (\theta_\delta + \theta_2 - \pi)/2$, where
$\theta_\delta$ is the phase of $\delta$.
\bibitem{footnoteM2} There is another solution 
when $m_2=0$ and $m_1\neq 0$ (non-symmetric
Gaussian states). Here the problem is solved by
exchanging the roles of Alice and Bob in the previous protocol:  
$m_2 \leftrightarrow m_1$ and $n_2 \leftrightarrow n_1$ in all expressions.
\end{thebibliography}
\end{document}